\documentclass[fleqn,10pt]{wlscirep}
\usepackage[utf8]{inputenc}
\usepackage[T1]{fontenc}

\usepackage{hyperref}
\usepackage{txfonts}
\usepackage{graphicx}

\usepackage{xspace}
\newcommand{\lcvo}{LiCuVO\ensuremath{_4}\xspace}
\newcommand{\tn}{\ensuremath{T_\mathrm{N}}\xspace}
\newcommand{\aaxis}{\ensuremath{a}\xspace}
\newcommand{\baxis}{\ensuremath{b}\xspace}
\newcommand{\caxis}{\ensuremath{c}\xspace}
\renewcommand\vec{\mathbf}
\usepackage{epstopdf}
\title{Domain dynamics in \lcvo: Evidence for polarized nanoregions}

\author[1,*]{Christoph P. Grams}
\author[1]{Severin Kopatz}
\author[1]{Daniel Br\"uning}
\author[1]{Sebastian Biesenkamp}
\author[2]{Petra Becker}
\author[2]{Ladislav Bohatý}
\author[1]{Thomas Lorenz}
\author[1]{Joachim Hemberger}

\affil[*]{grams@ph2.uni-koeln.de}
\affil[1]{University of Cologne, Institute of Physics II, Z\"ulpicher Str. 77, 50937 Cologne, Germany}
\affil[2]{University of Cologne, Institute of Geology and Mineralogy, Section Crystallography, Z\"ulpicher Str. 49b, 50674 Cologne, Germany}

\begin{abstract}
\lcvo is a model system of a 1D spin-1/2 chain that enters a planar spin-spiral ground state below its N\'eel temperature of 2.4\,K due to competing nearest and next nearest neighbor interactions.
The spin-spiral state is multiferroic with an electric polarization along the \aaxis axis which has been proposed to be caused purely by the spin supercurrent mechanism.
With external magnetic fields in \caxis direction \tn can be suppressed down to 0\,K at 7.4\,T.
Here we report dynamical measurements of the polarization from $P(E)$-hysteresis loops, magnetic field dependent pyro-current and non-linear dielectric spectroscopy as well as thermal expansion and magnetostriction measurements at very low temperatures.
The multiferroic transition is accompanied by strong anomalies in the thermal expansion and magnetostriction coefficients and we find slow switching times of electric domain reversal. Both observations suggest a sizable magnetoelastic coupling in \lcvo. 
By analyzing the non-linear polarization dynamics we derive domain sizes in the nm range that are probably caused by Li defects.

\end{abstract}

\begin{document}

\flushbottom
\maketitle
\thispagestyle{empty}

\section*{Introduction}
Multiferroic materials are in the focus of fundamental condensed matter research because of their complex physical properties arising from the coupled magnetic and ferroelectric order parameters~\cite{Cheong2007, Tokura2014}.  
Moreover, they also bear a huge potential for technological applications in, e.g., data storage or sensor technologies~\cite{Spaldin2005, Fiebig2016, Spaldin2017}.
Of particular importance in this context is a detailed understanding of the dynamics of domain switching processes. 
The material \lcvo originally sparked the scientific interest by showing a broad maximum in temperature dependent magnetic susceptibility~\cite{Blasse1966}, that has subsequently been shown to be caused by 1D spin 1/2 chains. 
The spin chains are composed of $S=1/2$ Cu$^{2+}$ ions that are coupled by competing ferromagnetic nearest ($J_1<0$) and antiferromagnetic next nearest neighbor ($J_2>0$) interactions along the crystallographic \baxis axis~\cite{Enderle2005}.
This frustration leads to the formation of multiferroic cycloidal spin order with a second-order N\'eel transition at 2.4\,K in zero magnetic field~\cite{Gibson2004,Mourigal2011}.
Contrary to many other multiferroic compounds, e.g. the manganites~\cite{Mochizuki2009}, MnWO$_4$~\cite{Lautenschlaeger1993}, or Ni$_3$V$_2$O$_8$~\cite{Chaudhury2007}, a collinear spin density wave phase above this phase has not been observed. 
The ferroelectric polarization $\vec{P_r} \propto \vec{k} \times (\vec{S_n} \times \vec{S_{n+1}})$ with $\vec{k}\parallel\baxis$ is parallel to the \aaxis axis and has been proposed to be driven purely by the spin supercurrent mechanism~\cite{Mourigal2011}.

High magnetic fields induce additional phase transitions~\cite{Buttgen2007,Svistov2011} at low temperatures.
With $H \parallel \caxis$ the electric polarization of the sample is suppressed as the spin structure realizes a collinear spin modulated phase above about 7.5\,T. 
At very high magnetic fields of about 41\,T, \lcvo enters a spin-nematic phase before the fully spin-saturated phase is reached above 44\,T~\cite{Svistov2011,Mourigal2012}.
 
The goal of our measurements was to study the switchability of the polarization in order to derive both the underlying coupling and domain switching mechanisms.
As it turns out, our results call the proposed spin supercurrent mechanism into question by demonstrating the presence of a sizable magnetoelastic coupling and slow dynamics of the ferroelectric domain switching.
Furthermore, the low growth dimension and the Vogel-Fulcher-like temperature dependence of the characteristic relaxation time are evidence for a distribution of domain sizes in the nm range determined by structural defects.


\section*{Results}

\subsection*{Crystal growth}
\begin{figure}
  \centering
  \includegraphics[width=0.6\columnwidth]{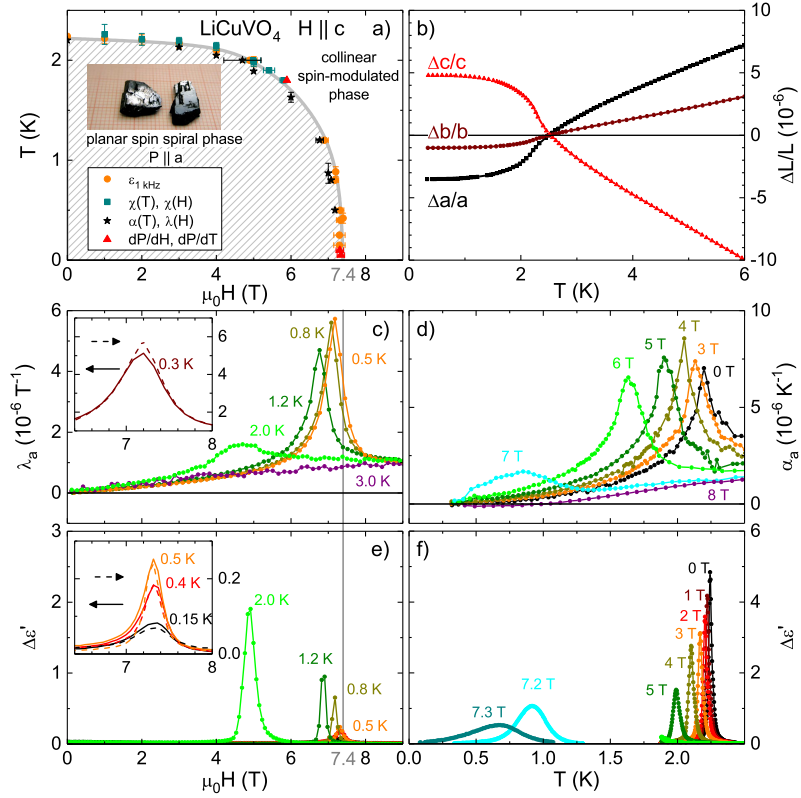}
  \caption{
    (a) Phase diagram of \lcvo with $H \parallel \caxis$.
    (b) Relative length change $\Delta L_i/L_i$ measured parallel to $i=\aaxis$,\baxis, and \caxis in $\mu_0 H=0$\,T . 
    Both the magnetostriction coefficient $\lambda_a$ in (c) and the thermal expansion $\alpha_a$ in (d) show the transition into the multiferroic phase as peaks.
    The phase transition is similarly observed in $\Delta\varepsilon'(H)$ in (e) and $\Delta\varepsilon'(T)$ in (f) measured at $\nu = 1$\,kHz.
    Inserts in (c) and (e) show enlarged views of the respective low-$T$ data measured either with increasing (dashed) or decreasing (solid) magnetic field.
  }
  \label{fig:phasediagram}
\end{figure}

\lcvo has an orthorhombically distorted inverse spinel structure of the space group \textit{Imma}\cite{Lafontaine1989} and, on heating, shows peritectic decomposition into CuO and a LiVO$_3$-rich melt. 
For the growth of single crystals there is an access path for compositions LiVO$_3$ - \lcvo from 10 to 60 mol\% \lcvo \cite{Prokofiev2000, Prokofiev2005}. 
We identified a starting composition of 35 mol\% \lcvo, a starting temperature of 910 K and a cooling rate of 0.1 K/h as suitable crystal-growth conditions.
Resulting crystals show a platy habit with well-developed morphological face (001) and dimensions up to 15 x 8 x 5 mm$^3$ as shown in the inset of Fig. \ref{fig:phasediagram}(a).

\subsection*{Phase diagram}

Figure~\ref{fig:phasediagram}(a) shows the temperature vs. magnetic-field phase diagram obtained from our measurements for a magnetic field applied along the \caxis axis. 
We restrict ourselves to $H \parallel \caxis$, because for this field direction only the transition from the multiferroic, ferroelectric phase to the paraelectric phase occurs, whereas additional spin-flop transitions are induced for $H \parallel \aaxis$ or $\baxis$~\cite{Schrettle2008}, which complicate the analysis of the domain dynamics. 
The multiferroic ordering transition causes significant anomalies in the thermal expansion coefficients $\alpha_i$ measured along all three lattice constants $i=a,b,c$. 
As is shown in Fig.~\ref{fig:phasediagram}(b), $\Delta a/a$ and $\Delta b/b$ spontaneously contract in the ordered phase and the sum of these contractions is essentially compensated by the expansion of $\Delta c/c$. 
Strongly anisotropic strains $\Delta L_i/L_i$ of comparable magnitudes have also been reported at the multiferroic ordering transitions of MnWO$_4$~\cite{Chaudhury2008}.
These strains reveal a pronounced magnetoelastic coupling in \lcvo, which result from significant uniaxial pressure dependencies of the exchange couplings and also cause the negative thermal expansion of the \caxis axis above $T_{\rm N}$. 
In order to derive the phase boundary we measured $\Delta L_i(H,T)$ along the \aaxis axis either as a function of $H \parallel c$ at constant $T$ or as a function of $T$ at constant $H$ and track the pronounced maximums of the magnetostriction $\lambda_a = 1/L_a  \partial \Delta L_a / \mu_0\partial  H$ and thermal expansion coefficients $\alpha_a = 1/L_a  \partial \Delta L_a / \partial  T$, which are shown in Figs.~\ref{fig:phasediagram}(c) and (d), respectively. 
This phase transition also causes corresponding peaks in the real part of the dielectric constant $\varepsilon(T,H)$ measured at a frequency of 1\,kHz, which are displayed in Figs.~\ref{fig:phasediagram}(e) and (f) after subtracting the background measured in the paraelectric phase, i.e. $\Delta \varepsilon' = \varepsilon' - \varepsilon_\infty$. 
The peak positions of the dielectric and of the expansion measurements are in good agreement, but the temperature- and field-dependent evolution of the peak shapes systematically differ.
For example, the peaks of $\lambda_a$ sharpen on decreasing temperature, whereas those of $\Delta\varepsilon'$ strongly decrease in magnitude; see panels~(c) and (e). 
Despite this systematic difference, the dielectric and the expansion measurements consistently reveal no indications of hysteresis between the data obtained with increasing or decreasing magnetic field, as is shown in the corresponding insets. 
This confirms that the nature of this phase transition remains of second order down to the lowest temperature of our measurement with the critical field $\mu_0H_{\rm c}=7.4$\,T at $50$\,mK.

\begin{figure} 
  \centering 
  \includegraphics[width=0.66\columnwidth]{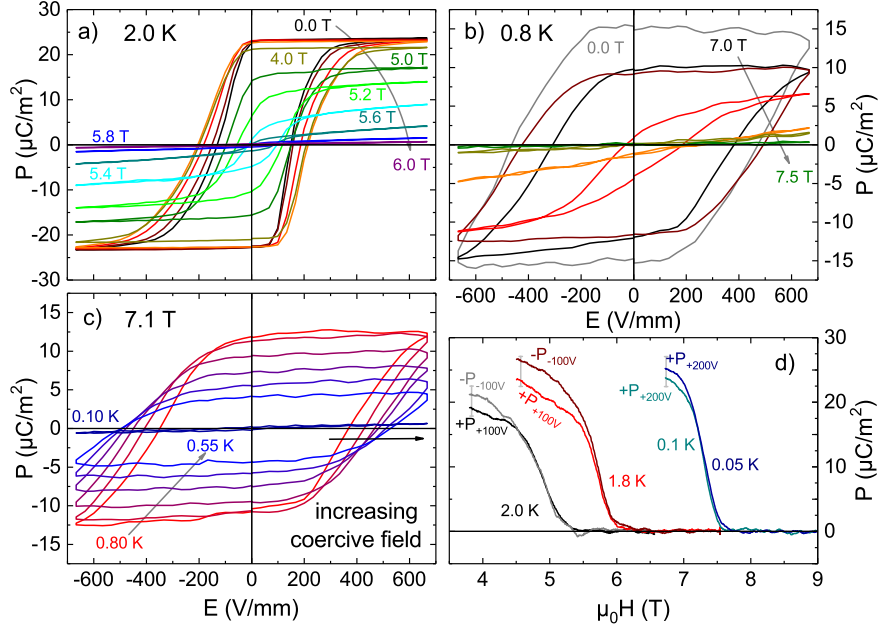}
  \caption{
    $H$-dependent $P(E)$ loops at 2.0\,K are shown in (a) and at 0.8\,K in (b).
    (c) shows $T$-dependent $P(E)$ loops at $\mu_0H={\rm 7.1\,T}$ below 0.8\,K.
    Magneto-current measurements in (d) show the multiferroic phase transition down to 0.05\,K.
  }
  \label{fig:polarization}
\end{figure}

\subsection*{Quasi-static polarization}

To evaluate the switchability of the polarization we measured quasi-static $P(E)$ hysteresis loops with an effective frequency $\nu_{P(E)} \approx 0.01{\rm \,Hz}$ that are shown in Fig. \ref{fig:polarization}. 
For better visibility the dielectric background $\varepsilon_\infty \approx 7.4$ was subtracted from all curves.
At 2.0\,K shown in Fig. \ref{fig:polarization}(a) the hysteresis curves for $\mu_0H \lesssim 4$\,T are fully formed and reach saturation, similar to measurements already published~\cite{Ruff2014,Ruff2017}.
With further increasing the magnetic field the loops start to close and above the critical field $\mu_0H_{\rm c}({\rm 2\,K}) = 5.8$\,T the expected paraelectric linear $P(E)$ is observed.
 
When cooling the sample to 0.8\,K (Fig. \ref{fig:polarization}(b)) it is no longer possible to switch the polarization with the applied electric fields.
Here, the hysteresis loops are only fully formed very close to the critical field $\mu_0 H_\mathrm{c}(0.8\,\mathrm{K})=7.3$\,T, demonstrating that the coercive field is higher compared to 2.0\,K.
This can be seen even better in the temperature dependence of $P(E)$ loops at 7.1\,T in Fig. \ref{fig:polarization}(c). 
For $0.8{\rm \,K} <T< 0.55{\rm \,K}$ each decrease of the temperature by 0.05\,K increases the coercive field and, finally, at 0.1\,K the coercive field so strongly exceeds the applied electric field that the $P(E)$ loop is essentially flat.

To observe the magnetic field dependence of the quasi-static spontaneous polarization also at lower temperatures where the high coercive fields prevent full polarization switching we performed magneto-current measurements.
The polarization $P(H)$, calculated by integration of $I_\mathrm{mag}(t) - I_\mathrm{leakage}$, is shown in Fig. \ref{fig:polarization}(d) where the spontaneous polarization at the multiferroic phase transition can be seen down to 0.05\,K.
Within the experimental accuracy the spontaneous polarization below 1.8\,K approaches an essentially constant value $P_{\rm S}(T \to 0) \approx 30$\,$\mu$C/m$^2$.

\subsection*{Polarization dynamics}

As an alternative to the quasi-static "dc" measurements of the polarization described above, we also use sinusoidal electric fields $E(t)=E_\mathrm{ac} \sin \omega t$.
In this case, the polarization $P(t)$ is obtained by expansion of the applied electric field, 

\begin{equation}
 P(t) = \varepsilon_0 E_\mathrm{ac}  \sum\limits_{n=1}^\infty (\varepsilon_n' \sin n\omega t - \varepsilon_n'' \cos n\omega t ).
\end{equation}

In the experiment, this approach is realized by measuring the higher harmonics of the excitation frequency with the lock-in technique which makes a much broader frequency range up to kHz accessible.
For square-type hysteresis loops these measurements in the frequency domain can be used to determine the coercive field $E_\mathrm{crcv}$ and the switchable polarization $P_\mathrm{sw}$ from the lowest-order components $\varepsilon_1'$ and $\varepsilon_1''$ because even terms vanish due to the inversion-symmetry of the ferroelectric hysteresis loops and the magnitude of higher-order terms decreases with $1/n$.
As discussed in detail in~\cite{Niermann2014} the linear contributions to $\varepsilon_1'$ from phonon modes, $\varepsilon_\infty$, have to be removed and both switchable polarization and coercive field can then be calculated with $P_\mathrm{sw} = \frac{\pi}{4} \varepsilon_0 E_\mathrm{ac} |\Delta\varepsilon_1|$ and $E_\mathrm{crcv} = E_\mathrm{ac} \varepsilon''_1/|\Delta\varepsilon_1|$ respectively, where $|\Delta\varepsilon_1| = ({(\varepsilon'_1 - \varepsilon_\infty)}^2 + {\varepsilon''_1}^2)^{1/2}$.

\begin{figure} 
  \centering
  \includegraphics[width=0.66\columnwidth]{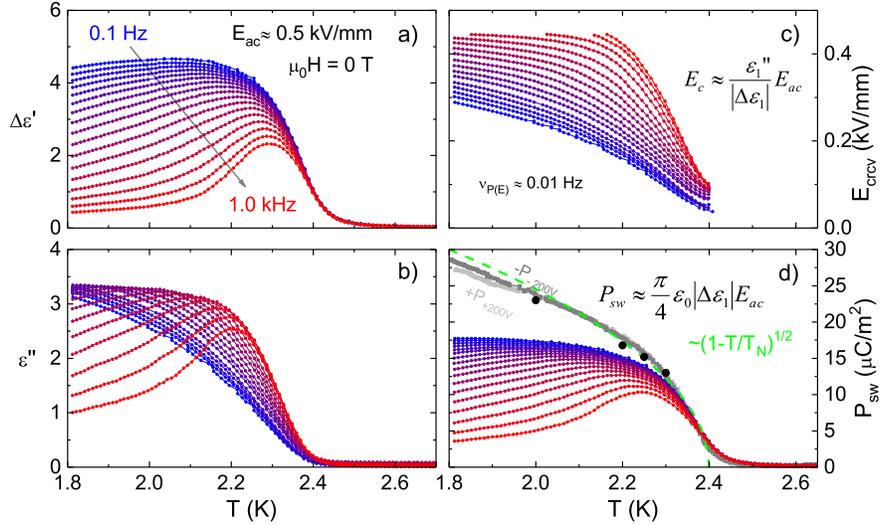}
  \caption{
  (a) and (b) show measurements of the real and imaginary part of the complex permittivity from 0.1\,Hz to 1\,kHz at $E_\mathrm{ac} = 500$\,V/mm in zero magnetic field.
  From this data we calculate $E_\mathrm{crcv}$ in (c) and $P_\mathrm{sw}$ in (d), the latter is compared to results from quasi-static $P(E)$ measurements (black) and pyrocurrent (gray).
  The green line in (d) show the $T$ dependence of the polarization expected from mean-field theory. 
  }
  \label{fig:hvbT}
\end{figure}

The complex first-order non-linear permittivity $\varepsilon_1^*$ was measured in zero magnetic field with $E_\mathrm{ac} = 0.5$\,kV/mm, see Fig. \ref{fig:hvbT}(a) and (b).
$\varepsilon_\infty$ has been removed from the real part of the permittivity by subtracting the measured results in the paraelectric phase at 3.0\,K.
In Fig. \ref{fig:hvbT}(c) we see the coercive field increasing with decreasing temperature; at the same time this increase is much steeper for higher frequencies where the curves are cut off when $E_\mathrm{crcv}$ reaches about 90\% of the applied electric field.
The switchable polarization $P_\mathrm{sw}(T,\nu)$ is shown in Fig. \ref{fig:hvbT}(d).
Here, the transition into the ferroelectric phase can be seen in the increase of $P_\mathrm{sw}$ at the Néel temperature $T_\mathrm{N}\approx2.4$\,K.
While $P_\mathrm{sw}(T,\nu)$ increases with decreasing temperature for low frequencies, at high frequencies a maximum appears when the coercive field approaches the applied electric field.
For comparison of the absolute values static measurements of the saturation polarization $P_\mathrm{s}$ extracted from pyrocurrent (gray) and $P(E)$ loops (black) are also shown.
Additionally, the dashed line is shown as a guide to the eye 
\begin{equation}
 P_\mathrm{s} \approx 60 \frac{\mu\mathrm{C}}{\mathrm{m}^{2}} (1-T/T_\mathrm{N})^{1/2}
 \label{eq:Ps}
\end{equation}
where the exponent $1/2$ is expected from mean-field theory at a continuous phase transition~\cite{Toledano2010} and describes the results from the pyrocurrent measurements very well down to $T \approx 1.8{\rm \,K}$.
As the switchable polarization even at lowest frequencies stays clearly below the Landau-type behavior extrapolated from static measurements, part of the sample's polarization is pinned.

\begin{figure} 
  \centering
  \includegraphics[width=0.66\columnwidth]{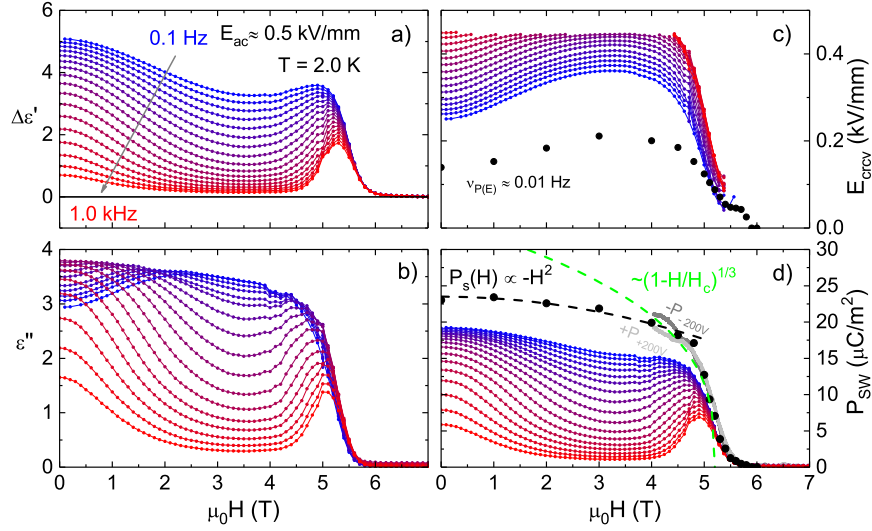}
  \caption{
  (a) and (b) show measurements of the real and imaginary part of the complex permittivity from 0.1\,Hz to 1\,kHz at $E_\mathrm{ac} = 0.5$\,kV/mm at 2.0\,K.
  From this data we calculate $E_\mathrm{crcv}$ in (c) and $P_\mathrm{sw}$ in (d), the latter is compared to results from quasi-static $P(E)$ measurements (black) and magneto-current (gray).
  The green line in (d) show the expected $H$ dependence of the polarization at the phase transition from mean-field theory as well as its $-H^2$ (black line) dependence in the multiferroic phase.
  }
  \label{fig:hvbH}
\end{figure}

Similar magnetic-field dependent measurements at 2.0\,K are shown in Fig. \ref{fig:hvbH}.
A key difference to the measurements in zero magnetic field is that with increasing magnetic field the spins are canted out of the $ab$-plane by a magnetic-field dependent canting angle $\theta$.
Thus, the saturation polarization along the \aaxis axis also depends on $H$ as $P_\mathrm{s} \propto \cos^2(\theta)$, i.e. the polarization in the multiferroic phase will be reduced with increasing magnetic field.
To quantify the magnetic field dependence further, we use the result from magnetization measurements in \caxis direction that show $M_\caxis \propto \sin(\theta) \propto H$ in the multiferroic phase~\cite{Svistov2011}.
This leads to an expected field dependence of $P_\mathrm{s}(H) - P_\mathrm{s}(0) \propto - H^2$ that agrees very well with our results from $P(E)$ measurements, see Fig. \ref{fig:hvbH}(d) (black line).
Due to this reduction of $P_\mathrm{s}$ and its interplay with the coercive field we have to distinguish two magnetic-field regimes in our frequency dependent measurements.
At $\mu_0H = 0$\,T we start in the multiferroic phase with frequency-dependent splitting of the curves as discussed above.
As the magnetic field is increased to $\mu_0H \approx 3$\,T the coercive field increases and, consequently, the switchable polarization is reduced.
Here, the depression in the switchable polarization demonstrates that $E_{\rm ac} = 0.5$\,kV/mm is insufficient to switch the non-pinned polarization fully for frequencies above 0.3\,Hz
Close to the phase transition at $\mu_0 H_\mathrm{c}(2.0$\,K$) \approx 5.2$\,T the coercive field drops down and $P_\mathrm{sw}$ increases.
Once the coercive field is low enough $P_\mathrm{sw}(H)$ has a maximum and then follows the saturation polarization that vanishes in the paraelectric phase.
In a field dependent measurement also the expected critical exponents at the phase transition differ from the temperature dependent behavior.
According to canonical Landau theory, a ferroelectric material with a linear magnetoelectric contribution to the free energy near the multiferroic phase transition obeys $P_\mathrm{s}\propto(1-H/H_\mathrm{c})^{\beta_H}$ with $\beta_H = 1/3$~\cite{Toledano2010,Kim2009}.
The quasi-static results follow this prediction down to $H \approx 0.8H_{\rm c}$ while the switchable polarization measured with the dynamical approach follows this prediction only in a much smaller magnetic-field range.

\subsection*{Discussion}

Due to the broad frequency and electric-field range available with the above method we can use the Ishibashi-Orihara model for growth- and nucleation-dominated domain switching~\cite{Orihara1994} to analyze the underlying mechanism for for the switching dynamics of the polarization.
This model parametrizes the switching dynamics via an effective domain-growth dimension $d$ and an exponent $\alpha$ correlating the switching process with the applied electric field.
For MnWO$_4$, such an analysis reveals a dimensionality $d_{\mathrm{MnWO}_4}=1.8$ and $\alpha_{\mathrm{MnWO}_4} = 3.6$~\cite{Niermann2014}.
From the magnetic structure of \lcvo one may expect $d \gtrsim 1$ reflecting the weakly coupled 1D spin chains.
A key prediction of the Ishibashi-Orihara model relates $P_\mathrm{sw}$ to the frequency $\nu$ and an electric-field dependent factor $\Phi_E$ via
\begin{equation}
 P_\mathrm{sw} \propto 1-\exp(-\nu^{-d}\Phi_E).
\end{equation}

\begin{figure} 
  \centering
  \includegraphics[width=0.66\columnwidth]{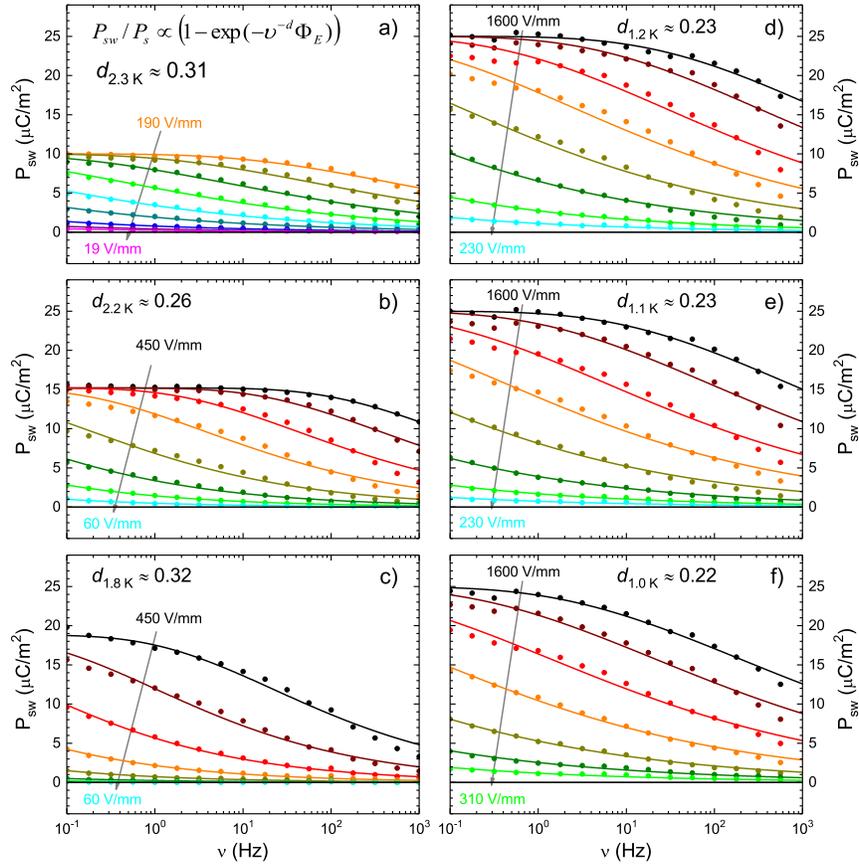}
  \caption{
    Switchable polarization (symbols) at different temperatures fitted via the Ishibashi-Orihara model (lines) yielding the effective dimensionality $d$ and a parameter $\Phi_E$.
  }
  \label{fig:ishibashi}
\end{figure} 

We derive $d$ and $\Phi_E$ from the corresponding fits of the $P_\mathrm{sw}(\nu)$ data in zero magnetic field at different temperatures. 
Fig. \ref{fig:ishibashi} displays the measured $P_\mathrm{sw}(\nu)$ together with fits for $T=$2.3, 2.2, 1.8, 1.2, 1.1, and 1.0 K.
At all temperatures the results are well described by this model and at all temperatures we find $d \ll 1$ with an average value of $d=0.26(4)$.
The second parameter in this model, $\Phi_E$, depends on the applied electric field $E_\mathrm{ac}$ and its field dependence can be described by
\begin{equation}
 \Phi_E = \Phi_0 \cdot \left(\frac{E_\mathrm{ac}}{E_\mathrm{crcv}}\right)^\alpha.
\end{equation}

\begin{figure} 
  \centering
  \includegraphics[width=0.66\columnwidth]{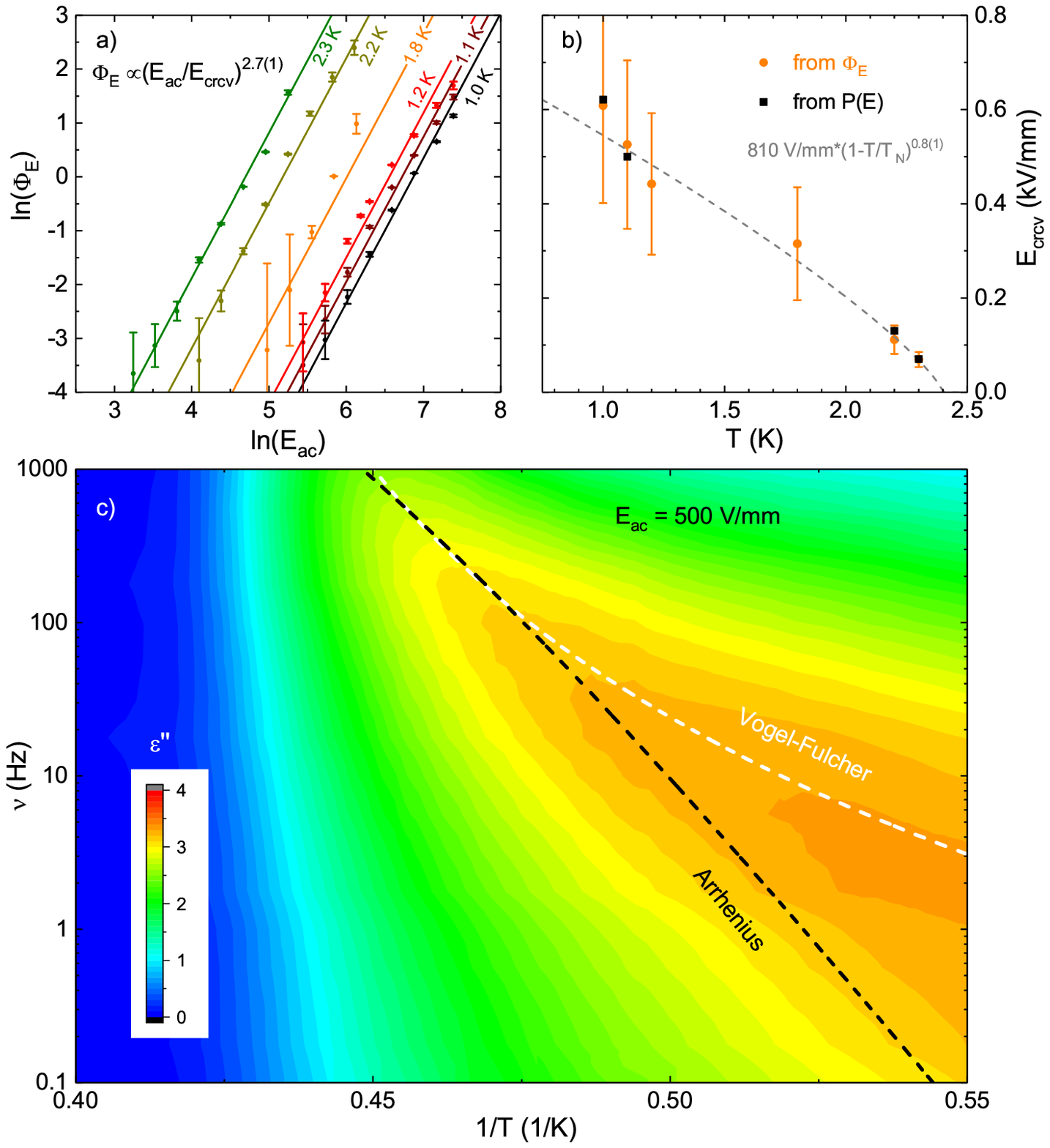}
  \caption{
    (a) The parameters $\Phi_E$ are described by $\Phi_E = \Phi_0 (E_\mathrm{ac}/E_\mathrm{crcv}(T))^\alpha$ with $\alpha=2.7(1)$ for all temperatures.
    In (b) the $T$-dependence of $E_\mathrm{crcv}$ is shown and can be described with $E_\mathrm{crcv}(T) \approx \mathrm{810\,V/mm}\cdot(1-T/T_N)^{0.8}$. 
    (c) $\varepsilon''(\nu, 1/T)$ with fits to the maximums of the measured spectra for Arrhenius-like (black) and Vogel-Fulcher-like (white) behavior.
  }
  \label{fig:ishibashi2}
\end{figure} 
As shown in Fig. \ref{fig:ishibashi2}(a) we find $\alpha=2.7(1)$ when fitting all five temperatures simultaneously.
The prefactor $\Phi_0 \approx 0.3$ follows from a comparison to the coercive field seen in $P(E)$ loops at different temperatures.
In Fig. \ref{fig:ishibashi2}(b) the $T$-dependence of $E_\mathrm{crcv}$ is fitted assuming a power law similar to the saturation polarization, the result is
\begin{equation}
  E_\mathrm{crcv} (T) \approx 810 \frac{\mathrm{V}}{\mathrm{mm}} \cdot \left(1-\frac{T}{T_\mathrm{N}}\right)^{0.8}. 
  \label{eq:Ecrcv}
\end{equation}

On the one hand, our data can be almost perfectly parametrized by the model of Ishibashi and Orihara and, therefore, seems to correspond to other multiferroics like e.g. MnWO$_4$~\cite{Niermann2014}. 
On the other hand, however, the outcome of an effective growth dimension $d \ll 1$ suggests, that the switching process is not dominated by the growth of domains~\cite{Orihara1994}. 


A "point-like" growth dimension indicates the flipping of stiff domains, a process that induces thermally activated behavior with an energy barrier $U_\mathrm{B} \propto V_\mathrm{d} P_\mathrm{s}(T) E_\mathrm{crcv}(T)$ with $V_{\rm d}$ denoting the domain volume. 
Taking into account the above results from equations (\ref{eq:Ps}) and (\ref{eq:Ecrcv}) we find a temperature dependent energy barrier $U_\mathrm{B}$ as 

\begin{equation}
U_\mathrm{B}(T)  \propto V_\mathrm{d} P_\mathrm{s}(T) E_\mathrm{crcv}(T) \approx V_\mathrm{d} \cdot 49 \frac{\mathrm{J}}{\mathrm{m}^{3}} \cdot \left( 1-\frac{T}{T_\mathrm{N}}\right)^{1.3}.
\end{equation}


Figure \ref{fig:ishibashi2}(c) shows the $T$ and $\nu$ dependence of the dielectric loss $\varepsilon''(T,\nu)$. 
At each temperature the dielectric loss spectra show a maximum at the frequency $\nu_{\rm p}$ which corresponds to the relaxation time for switching domains of predominant size.
This in turn is compared to an Arrhenius-like (black) and a Vogel-Fulcher-like (white) temperature dependence. 
The Arrhenius approach $\nu_{\rm p}(T) \propto \exp(-U_\mathrm{B}/k_\mathrm{B}T)$ only works close to the phase transition and the increase of $U_\mathrm{B}$ at lower temperatures would increase the slope of the curve bending it further away from the measured data. 
In contrast, a Vogel-Fulcher-like approach $\nu_{\rm p}(T) \propto \exp(U_\mathrm{B}/(T-T_\mathrm{VF}))$ describes the temperature dependency of the data over a much broader temperature range.
The additional parameter in this model, the Vogel-Fulcher temperature $T_\mathrm{VF}$, is found to be $T_\mathrm{VF} \approx T_\mathrm{N}$.
While this type of temperature dependence is typically discussed in the context of glass-like dynamics it has also been used to model the dynamics of polar nanoregions in relaxor ferroelectrics~\cite{Pirc2007}.
Despite their different underlying microscopic models both fits yield similar results for the domain volume; $V_\mathrm{d,A} \approx 3.8\cdot10^{-23}$\,m$^3$ and $V_\mathrm{d,VF} \approx 2.0\cdot10^{-23}$\,m$^3$ which correspond roughly to spheres with a diameter of 38\,nm.
This value of course has to be understood as an average over a wide distribution of domain sizes and corresponding switching probabilities as is also suggested by the large widths of the spectra shown in Fig. \ref{fig:ishibashi}.

It has been reported that even carefully prepared \lcvo crystals contain a few percent of Li defects~\cite{Prokofiev2004}.
Although the Li in \lcvo is nonmagnetic it has been argued that the resulting hole-doped oxygen sites can form singlets with Cu spins that are equivalent to nonmagnetic defects on Cu sites~\cite{Prokofiev2004,Zhang1988}.
The influence of such defects in \lcvo has also been discussed in the context of NMR~\cite{Buttgen2014} and specific heat measurements~\cite{Prozorova2015}. 
As is argued in~\cite{Prozorova2015}, non-magnetic defects hardly disturb the formation of the spin-spiral state because of the next-nearest neighbor (NNN) exchange $J_2$ which still favors the antiparallel alignment of NNN spins that are separated by a nonmagnetic defect. 
However, such defects are natural sources for antiphase domain boundaries where the spin spiral changes from clockwise to anticlockwise and, accordingly the electric polarization switches from up to down. 
Concerning the spin-modulated phase, however, non-magnetic defects act as random phase shifts~\cite{Buttgen2014}, which severely disturb the formation of long-range incommensurate order and can explain why our thermal expansion data at fields above 7.5\,T do not show sizable anomalies that signal a sharp phase boundary between the paramagnetic and the spin-modulated phase. 
A very similar situation is present in the effective Ising spin-1/2 chain material BaCo$_2$V$_2$O$_8$: large thermal expansion anomalies signal a commensurate spin ordering, whereas a magnetic-field induced incommensurate spin-ordering is almost invisible in thermal expansion~\cite{Niesen2013}.   

In summary, by using high quality single crystalline samples we were able to extend the phase diagram of \lcvo with $H \parallel \caxis$ down to 0.05\,K with measurements of permittivity and magneto-current.
Our thermal expansion and magnetostriction data show clear anomalies at the phase transition of the multiferroic phase and demonstrating a sizable magnetoelastic coupling.
This is compatible with the slow switchability and polarization dynamics from which we conclude that the polarization in \lcvo is strongly pinned.
Our analysis rules out a domain-growth dominated switching process, instead the polarization dynamics seems to be determined by a distribution of fixed domain sizes in the nm range.
We presume that the domain sizes are determined by Li defects that in turn cause the Cu to form nonmagnetic singlets with hole doped oxygen sides.

\section*{Methods}

For the polarization measurements two contacts where applied with silver paint on opposing ends of the sample along the \aaxis axis.
The resulting capacitive signal was measured with a Novocontrol Alpha-A Analyzer at frequencies up to 1\,kHz in high ac electric fields, additional magneto- and pyro-current measurements where performed with a Keithley electrometer 6517B.
The dielectric measurements where done in two cryostats, a Quantum Design PPMS and a top-loading dilution refrigerator (Oxford Instruments KELVINOX).

High-resolution measurements of the relative length changes $\Delta L(T,H)/L$ were performed in a home-built capacitance dilatometer that was attached to a $^3$He system (Oxford
Instruments Heliox). The corresponding magnetostriction ($\lambda$) and thermal expansion ($\alpha$) coefficients were then obtained via numerical differentiation $(\alpha, \lambda) = \frac{1}{L_0}\frac{\partial \Delta L}{\partial (T,\mu_0H)}$.
 
\section*{Acknowledgements}

This work is funded through the Institutional Strategy of the University of Cologne within the German Excellence Initiative and the Deutsche Forschungsgemeinschaft via HE-3219/2-1 and CRC1238 (Projects A02, B01, and B02).

\section*{Author Information}

\subsection*{Contributions}

C.P.G., T.L., and J.H. planned and optimized the experiments, 
P.B. and L.B. grew the samples, 
C.P.G. and S.B. measured the polarization and its dynamics,
C.P.G. analyzed the polarization data and performed the calculations, 
D.B. and S.K. measured and evaluated the thermal expansion and magnetostriction,
C.P.G., T.L. and J.H. interpreted the experimental data and prepared the manuscript,
All authors discussed the results and commented on the manuscript.

\subsection*{Competing Interests}

The authors declare no competing interests.

\subsection*{Data availability}
The datasets measured and analysed during the current study are available from the corresponding author on reasonable request.

\end{document}